\documentstyle[preprint,aps,epsf,psfig]{revtex}
\def\beq{\begin{equation}}
\def\eeq{\end{equation}}
\def\barr#1{\begin{array}{#1}}
\def\earr{\end{array}}
\def\beqar{\begin{eqnarray}}
\def\eeqar{\end{eqnarray}}
\def\beqars{\begin{eqnarray*}}
\def\eeqars{\end{eqnarray*}}
\def\bitem{\begin{itemize}}
\def\eitem{\end{itemize}}

\begin{document}
\draft
\preprint{\begin{tabular}{l}
\hbox to\hsize{July, 1998 \hfill KAIST-14/98}\\[-3mm]
\hbox to\hsize{hep-ph/9807496 \hfill SNUTP 98-079}\\[5mm] \end{tabular} }
\vspace{1cm}
\title{A new method for extracting  the weak \\  phase $\gamma$     from 
$B \rightarrow D K^{(*)}$ decays}

\vspace{2cm}

\author{
Ji-Ho Jang, \footnote{e-mail~:~jhjang@chep6.kaist.ac.kr}
and Pyungwon Ko  \footnote{e-mail~:~pko@charm.kaist.ac.kr} }
\address{
Department of Physics, Korea Advanced Institute of Science and Technology,\\
Taejon 305-701, Korea}
\maketitle
\begin{abstract}
A new method to extract the weak phase $\gamma$ is suggested by exploiting
$B \rightarrow D K^{(*)}$ decay modes that are not Cabibbo  suppressed,
using the isospin relations, and ignoring the annihilation diagram as usual. 
Assuming $3 \times 10^8 B\bar{B}$ pair at $B$ factories, one can determine 
$\gamma$ with $3-\sigma$ accuracy for $80^o \lesssim \gamma \lesssim 150^o$ 
using $B \rightarrow D K$ modes and 
for $50^o \lesssim \gamma \lesssim 170^o$ using $B \rightarrow D K^*$ modes.   
\end{abstract}


\newpage
\narrowtext
\tighten

One of the goals of B factories which will launch its mission at the
end of 1999 is to test the KM paradigm for the CP violation by verifying
that the unitary triangle (UT) can be constructed in a consistent manner
\cite{quinn}.
Namely, one measures three sides and three angles of the unitarity triangle
in all possible ways, and examine if a single triangle emerges from
various different measurements. If so, one can verify that CP violations in 
$K_L \rightarrow \pi\pi$ and $B$ decays all result from a single KM phase
in the CKM matrix \cite{km}. 
Otherwise, there should be some new physics which is more 
exciting from the particle physics point of view. Still, it is utmostly 
important to test the KM picture from $B$ decays. As of today, the least 
known quantities of the UT are $|V_{ub}/V_{cb}|$, and its three angles 
\cite{aliburas} :
\begin{eqnarray}
| V_{ub}/V_{cb} | & = & 0.080 \pm 0.020,
\\
-1.0 \le &\sin 2\alpha& ~\le 1.0, 
\\
0.30 \le &\sin 2\beta& ~\le 0.88,
\\
0.27 \le &\sin^2 \gamma& ~\le 1.0.
\end{eqnarray}
The first quantity is determined from charmless semileptonic $B$ decays (both
inclusive and exclusive), and suffers from intrinsic theoretical 
uncertainties such as breakdown of the heavy quark mass expansion near 
the phase space boundary, or the poorly known $B\rightarrow \pi~ ({\rm or}~ 
\rho)$ semileptonic form factors. One can estimate the uncertainty in 
$|V_{ub}/V_{cb}|$ as $\sim 25\%$ conservatively. Three angles of the UT can be
loosely bounded from various low energy phenomenology. The angle $\beta( 
\equiv \phi_1)$ can be measured in the gold-plated mode, $B_d \rightarrow 
J/\psi K_S$ without any hadronic uncertainty. The angle $\alpha (\equiv 
\phi_2)$ can be measured from $B \rightarrow \pi \pi$, but there is some 
penguin contamination  that cannot be too small considering the recent 
observation of $B\rightarrow K\pi$ at the level of branching ratio of 
$ \sim 1.4 \times 10^{-5}$ by CLEO Collaboration \cite{cleo}. Still, one can 
hope to perform the isospin analysis and  remove the penguin contribution, 
thereby being able to extract the $\alpha (\phi_2)$ with a reasonable 
accuracy \cite{GL90}. 
The most difficult to measure is the angle $\gamma ( \equiv \phi_3 )$.
There have been a lot of suggestions and discussions about how to measure
this quantity at $B$ factories (running at the top of  $\Upsilon (4S)$ 
resonance) \cite{gamma,dunietz}. 
Unfortunately there is no best way to determine $\gamma$ from 
$B$ decays up to now. Any method suggested so far has some weak points,
e.g., involving measurements of decay modes that have too low branching 
ratios.  In Ref.~\cite{gamma}, the authors proposed to extract  $\gamma$ 
using the independent measurements of $B \rightarrow D^0 K, B \rightarrow 
\bar{D}^0 K$ and $B \rightarrow D_{CP} K$. 
However the charged $B$ meson decay mode $B^- \rightarrow \bar{D}^0 K^-$ 
is experimentally difficult to measure.
The reason is that the final $\bar{D}^0$ meson should be identified using 
$\bar{D}^0 \rightarrow K^+ \pi^-$, but it is difficult to distinguish it
from doubly Cabibbo suppressed  $D^0 \rightarrow K^+ \pi^-$ following color 
and CKM allowed $B^- \rightarrow D^0 K^-$. There are some variant methods 
to overcome these difficulties. In Ref.~\cite{soni}, Atwood {\it et al.} used 
different final states into which the neutral $D$ meson decays  to extract 
information of $\gamma$. In Ref.\cite{gronau98}, 
Gronau proposed that the angle $\gamma$ is determined
only using the color allowed decay modes, $B^- \rightarrow D^0 K^-,
B^- \rightarrow D_{CP} K^-$ and their charge conjugation modes.

In this letter, we suggest another method for extracting $\gamma$ from Cabibbo
allowed $B\rightarrow D K^{(*)}$ decays.  We construct three different 
triangles from various $B \rightarrow D K^{(*)} $ decays, each of which 
involves decay modes with rather large branching ratios. From these triangles,
one can determine the weak phase $\gamma$ with a reasonable accuracy if one 
has $3\times ~10^8 B \bar{B}$'s at $B$ factories.  
Both $B\rightarrow D K$ modes and 
the self-tagging modes $B\rightarrow D K^*$ are considered with an assumption
that the annihilation diagrams are negligible in both cases. This assumption
may be questionable for the $B \rightarrow D K^*$ decays in light of the 
recent work by Ali {\it et al.}  \cite{alilu}, 
which claims that the annihilation diagram 
may not be ignorable in the $B \rightarrow  P V$ channel for the case of 
light pseudoscalar ($P$) and light vector ($V$) mesons. This claim is based 
on the generalized factorization approximation. We leave this as an open 
question here, with a remark that one can easily test this assumption
by measuring the branching ratio for $B^- \rightarrow D^- \bar{K}^{*0}$
and comparing it with other decays we use, such as $B^{-(0)} \rightarrow D_1 
\bar{K}^{*-(0)}$ and $\bar{B^0} \rightarrow D_1 \bar{K}^{*0}$.

The recent CLEO measurement of $Br (B^+ \rightarrow \bar{D}^0 K^-) = (2.57
\pm 0.65 \pm 0.32) \times 10^{-4}$  \cite{cleo98} gives
light on the determination of one angle $\gamma$ of unitary triangle.
Let us begin with $B\rightarrow  D K$ 
and  define their amplitudes as follows, 
\begin{equation}
\label{amplitude1}
\begin{array}{lll}
A( B^- \rightarrow D^0 K^- )&= A \lambda^3~ B_1 e^{i \delta_1}&=
A \lambda^3~ ( T + C ) \\ \nonumber
A( \bar{B}^0 \rightarrow D^+ K^- )&= \frac{1}{2} A \lambda^3~ 
( B_1 e^{i \delta_1}- B_0 e^{i \delta_0}) &= A \lambda^3 ~ T  \\ \nonumber
A( \bar{B}^0 \rightarrow D^0 \bar{K}^{0} )&= \frac{1}{2} A \lambda^3 
~( B_1 e^{i \delta_1}+ B_0 e^{i \delta_0}) &= A \lambda^3 ~ C  \\ \nonumber
A( B^- \rightarrow \bar{D}^0 K^{-} )&= \frac{1}{2} A \lambda^3 
R_b e^{-i\gamma}~ ( B'_1 e^{i \delta'_1}+ B'_0 e^{i \delta'_0}) &= 
A \lambda^3 R_b e^{-i\gamma} ~( C' + A' )
\\ \nonumber
A( B^- \rightarrow D^- \bar{K}^{0} )&= \frac{1}{2} A \lambda^3 
R_b e^{-i\gamma}~ ( B'_1 e^{i \delta'_1}- B'_0 e^{i \delta'_0}) &= 
A \lambda^3 R_b e^{-i\gamma} ~( - A' )
\\ \nonumber
A( \bar{B}^0 \rightarrow \bar{D}^0 \bar{K}^{0} )&=  A \lambda^3 
R_b e^{-i\gamma} ~ B'_1 e^{i \delta'_1} &= A \lambda^3 R_b e^{-i\gamma}~  C'.
\end{array}
\end{equation} 
where we use the Wolfenstein parametrization of CKM matrix elements, and 
$R_b \equiv \sqrt{ \rho^2 + \eta^2 }$.  
$B_I^{(')}$ denotes the amplitude for the isospin $I$ state.
The first equalities are written in terms of  the isospin amplitudes, whereas 
the second ones are written in terms of diagramatic representations ($T$ 
means a tree diagram, $C$ means a color-suppressed diagram, and so on).
The above equations give two isospin relations,
\beqar
\label{isospin1}
A( B^- \rightarrow D^0 K^{-} )&=& ~~A( \bar{B}^0 \rightarrow D^+ K^{-} )
+ A( \bar{B}^0 \rightarrow D^0 \bar{K}^{0} )
\\ \nonumber
A( B^- \rightarrow \bar{D}^0 K^{-} )&=& - A( B^- \rightarrow D^- \bar{K}^{0} )
+ A( \bar{B}^0 \rightarrow \bar{D}^0 \bar{K}^{0} ).
\eeqar

Using the definition of mass eigenstates of $D$ mesons,
$D_{1(2)}=\frac{1}{\sqrt{2}} ( D^0 \pm \bar{D}^0 )$, and neglecting
the $D^0 - \bar{D}^0$ mixing, we obtain
\beqar
A( B^- \rightarrow D_1 K^{-} ) &=& A( \bar{B}^0 \rightarrow D_1 \bar{K}^{0} )
+ \frac{1}{\sqrt{2}} \left[ A( \bar{B}^0 \rightarrow D^+ K^{-} )-
A( B^- \rightarrow D^- \bar{K}^{0} ) \right] 
\\ \nonumber
A( B^- \rightarrow D_2 K^{-} ) &=& A( \bar{B}^0 \rightarrow D_2 \bar{K}^{0} )
+ \frac{1}{\sqrt{2}} \left[ A( \bar{B}^0 \rightarrow D^+ K^{-} )+
A( B^- \rightarrow D^- \bar{K}^{0} ) \right]. 
\eeqar

If we neglect  $A( B^- \rightarrow D^- \bar{K}^{0} )$ which is
CKM suppressed and has only the annihilation diagram contribution, we get
( from now, we only consider final $D_1$ states )
\beqar
\label{isospin2}
A( B^- \rightarrow D_1 K^{-} ) &=& A( \bar{B}^0 \rightarrow D_1 \bar{K}^{0} )
+ \frac{1}{\sqrt{2}}  A( \bar{B}^0 \rightarrow D^+ K^{-} )
\\ \nonumber
A( B^+ \rightarrow D_1 K^{+} ) &=& A( B^0 \rightarrow D_1 K^{0} )
+ \frac{1}{\sqrt{2}}  A( B^0 \rightarrow D^- K^{+} ).
\eeqar
The mixing of the neutral $B$ meson has to be considered in order to
obtain the magnitudes of $ A( B^0 \rightarrow D_1 K^{0} )$
and $A( \bar{B}^0 \rightarrow D_1 \bar{K}^{0} )$.
The time dependent decay rate of neutral $B$ mesons 
whose initial states are $B^0$ and $\bar{B}^0$ are given by
\beqar
\label{timerate}
\Gamma(B^0_{phys}(t) \rightarrow f)&=&
\frac{1}{2} |A_f|^2 e^{-\Gamma t} 
\left[ (1+|\xi|^2) +(1-|\xi|^2)\cos (\Delta m ~t)
-2 ({\mbox Im} \xi) \sin(\Delta m ~t ) \right],
\\ \nonumber
\Gamma(\bar{B}^0_{phys}(t) \rightarrow f)&=&
\frac{1}{2} |A_f|^2 e^{-\Gamma t} 
\left[ (1+|\xi|^2) -(1-|\xi|^2)\cos (\Delta m ~t)
+2 ({\mbox Im} \xi) \sin(\Delta m ~t ) \right].
\eeqar
Here, $\xi $ is 
\beq
\xi = e^{-2 i \phi_M } \frac{\bar{A}_f}{A_f},
\eeq
where $A \equiv A(B^0 \rightarrow f)$,
$\bar{A} \equiv A(\bar{B}^0 \rightarrow f)$,
$f = D_{1(2)} K_S$, and $\phi_M$ is the $B^0 - \bar{B}^0$ mixing phase.
It is possible to get 
$|A( \bar{B}^0 \rightarrow D_1 \bar{K}^{0} )|$ and 
$|A( B^0 \rightarrow D_1 K^{0} )|$ from the coefficients of the
constant and $\cos (\Delta m ~t)$ terms and using 
$|A( B^0 \rightarrow D_{1(2)} K^{0} )|=
|1/\sqrt{2} A( B^0 \rightarrow D_{1(2)} K_S )|$ and
$|A( \bar{B}^0 \rightarrow D_{1(2)} \bar{K}^{0} )|=
|1/\sqrt{2} A( \bar{B}^0 \rightarrow D_{1(2)} K_S )|$.

From Eq.~(\ref{amplitude1}), we get other relations for
$A(B^- \rightarrow D_1 K^{-})$ and its charge conjugate,
\beqar
\label{amplitude2}
A( B^- \rightarrow D_1 K^{-} ) &=& \frac{1}{\sqrt{2}}
A( B^- \rightarrow D^0 K^{-} ) + \frac{A \lambda^3 R_b}{2 \sqrt{2}}
B e^{i ( \delta - \gamma)} \\ \nonumber
A( B^+ \rightarrow D_1 K^{+} ) &=& \frac{1}{\sqrt{2}}
A( B^- \rightarrow D^0 K^{-} ) + \frac{A \lambda^3 R_b}{2 \sqrt{2}}
B e^{i ( \delta + \gamma)}
\eeqar
where $B$ is given by
$B e^{i \delta} \equiv  B'_1 e^{i \delta'_1} + B'_0 e^{i \delta'_0}$
and the last terms in Eq.(\ref{amplitude2}) are only
$\frac{1}{\sqrt{2}} A(B^- \rightarrow \bar{D}^0 K^-)$ and
$\frac{1}{\sqrt{2}} A(B^+ \rightarrow D^0 K^+)$.

The strategy to determine $\gamma$ is as follows :
\bitem
\item Using the first equation of Eq.(\ref{isospin1}),
we can draw a triangle and fix 
$ \frac{1}{\sqrt{2}} A( B^- \rightarrow D^0 K^{-})$.
The bottom side of the triangle is 
$\frac{1}{\sqrt{2}} |A(\bar{B}^0 \rightarrow D^+ K^{-})| =
\frac{1}{\sqrt{2}} |A(B^0 \rightarrow D^- K^{+})| $.

\item Using Eq.(\ref{isospin2}), we can draw
two triangles whose  bottom side is
$\frac{1}{\sqrt{2}} |A(\bar{B}^0 \rightarrow D^+ K^{-})| =
\frac{1}{\sqrt{2}} |A(B^0 \rightarrow D^- K^{+})| $ and fix
$A(B^- \rightarrow D_1 K^{-})$ and $A(B^+ \rightarrow D_1 K^{+})$.

\item Using three fixed amplitudes,
$ \frac{1}{\sqrt{2}} A( B^- \rightarrow D^0 K^{-})$,
$A(B^- \rightarrow D_1 K^{-})$ and $A(B^+ \rightarrow D_1 K^{+})$,
the angle $2 \gamma$  is determined by Eq.(\ref{amplitude2}) up to some
discrete ambiguities (see Fig.~1).
\eitem


In Fig.~1, we show three triangles that can be constructed from our strategy.
The thick solid sides are exactly what were problematic in the 
GLW method \cite{gamma}, since it is almost impossible to experimentally 
measure those sides. In our case, we use only $B$ decay modes with 
relatively large branching ratios so that one can avoid the difficulties 
encountered in the GLW method.
The question still remains whether one can extract $2 \gamma$ from Fig.~1
with a reasonable accuracy by measuring various sides of three triangles, 
which we would like to address in the following.  


Let us estimate the uncertainty in the determination of the weak phase 
$\gamma$ by the method suggested in this letter, assuming that 
$3 \times 10^8$ and $10^{11}$ $B \bar{B}$'s  are available at the 
$\Upsilon (4S)$ resonance ($B$ factories using $e^+ e^-$ annihilation) 
and  hadron colliders (such as BTeV or LHCB), respectively. 
The observed number of events for each mode is 
\[
N_{\rm obs} = N_{\rm tot} \times Br \times f_{} \times \epsilon, 
\]
where $N_{\rm tot}, Br, f$ and $\epsilon$ are the 
total number of $B - \bar{B}$ events, branching ratios, 
observation fractions and detector efficiencies,  respectively.
From $N_{\rm obs}$ one can determine the uncertainty 
$\Delta N_{\rm obs}$ of the branching ratio. 
The $K^0$ is identified by the $K_S \rightarrow \pi^+ \pi^-$ mode 
and using the fact that the half of $K_S$ is $K^0$. 
The $K^{*0}, \bar{K}^{*0}$ mesons are distinguished using
$K^{*0} \rightarrow K^+ \pi^-$ and $\bar{K}^{*0} \rightarrow
K^- \pi^+$ modes and they are also used for self-tagging of $B^0$ and
$\bar{B}^0$ respectively. In $D^0, \bar{D}^0$ meson
tagging, we add the $D^0 \rightarrow K^- \pi^+ \pi^0$ mode to
the $D^0 \rightarrow K^- \pi^+$, 
$D^0 \rightarrow K^- \pi^+ \pi^+ \pi^-$ modes to increase the
observation rate. 
In Table~1, the tagging modes, observation rate and
detector efficiencies are summarized. For each collider, we assume 
the same detection efficiencies as Dunietz's work \cite{dunietz}. From
the experimental value for $Br (B^+ \rightarrow D^0 K^-) 
\sim 2.6 \times 10^{-4}$, one can extract the size of $|T+C|$.  
Assuming that $|C/T| \approx |C^{'}/T| \approx \lambda = 0.22$ as in Ref.~
\cite{rosner}, and allowing $C$ and $C^{'}$ to have phases $\delta_C$ and 
$\delta_{C^{'}}$ relative to the $T$ amplitude, one can estimate the branching 
ratios of other decay modes that participate in the triangles shown in Fig.~1.
Then the uncertainty of the amplitude $A = \sqrt{Br}$ is determined by 
$\Delta A/A = (1/2) (\Delta Br/Br) \approx 1/( 2 \sqrt{ N_{\rm obs}} )$. 
Using this information, we investigate the possibility of the determination
of $\gamma$ from three triangles and its uncertainty.

The results are shown in Fig.~2~(a) and (b), where the horizontal and the 
vertical axes represent $\gamma$ and $\Delta \gamma$ (in degrees),
respectively.  Since the phases $\delta_C$ and $\delta_{C^{'}}$ are unknown,
we considered four different cases with a fixed $\delta_C = 10^o$ : 
(i) $\delta_{C^{'}} = \delta_C$ (the real curve), (ii) $\delta_{C^{'}} = 
-2 \delta_C$ (the dashed curve), (iii) $\delta_{C^{'}} = 0$ (the dotted curve),
and (iv) $\delta_{C^{'}} = +2 \delta_C$ (the dashed-dotted curve).  We observe
some dependence of  $\Delta \gamma$ on the phases $\delta_C$ and 
$\delta_{C^{'}}$ through the branching ratios.  Our method can provide a 
good determination of $\gamma$ for $80^{\circ} \lesssim \gamma \lesssim 
150^{\circ}$ or so. One can achieve an accuracy of better than 3-$\sigma$ for 
this range of $\gamma$. For small $\gamma$, our method fails and one has to 
resort to  other methods.

%

%
%

Now, let us repeat the same analysis for  the self-tagging modes, 
$B\rightarrow D K^*$.  The advantage of these self-tagging modes is that the 
number of available $B$ decays become doubled compared to the $B\rightarrow  
D K$ modes and the time dependent analysis is  unnecessary.
As before, we can define several amplitudes  similarly
to Eqs.~(\ref{amplitude1}) by $K \rightarrow K^*$. 
Then the same equations as (\ref{isospin1}),(\ref{isospin2}) and 
(\ref{amplitude2}) can be obtained. 
One thing to be kept in mind is the adequacy of neglecting
the amplitude for $B^- \rightarrow D^- \bar{K}^{*0}$ and its charge 
conjugate, which are Cabibbo suppressed and generated by the annihilation 
diagram at the quark level. Usually such annihilation diagrams are neglected,
since they are suppressed by $f_B / m_B$ relative to other diagrams. 
However this may not be true for the case of $B\rightarrow PV$ modes as 
recently discussed by Ali {\it et al.} in  the context of the 
generalized factorization approach \cite{alilu}.  
They claim that the annihilation branching ratio might be an order of 
magnitude higher than that of the penguin diagram only.
One can verify the usual assumption of neglecting the annihilation diagram
in $B\rightarrow D K^*$ only through the experimental measurement of the 
branching ratio for $B^- \rightarrow D^- \bar{K}^{*0}$. 

With this point kept in mind, we may proceed as before to construct three
triangles as shown in Fig.~1, and determine $\gamma$. For the estimate of the
uncertainties, let us assume that the branching ratio for 
$B^0 \rightarrow D^- K^{*+}$
is about $4\times 10^{-4}$ adopting the results of Neubert and Stech based on
the factorization approximation for $B$ decays into heavy and light mesons
\cite{neubert}.  Then, assuming the same relation between $T,C$ and $C^{'}$ and
their relative phases as before ($B\rightarrow D K$), one can estimate the 
uncertainties in the sides of three triangles in Fig.~1 (with $K\rightarrow 
K^*$) and the weak phase $\gamma$.  The results are shown in Fig.~2~(c) 
and (d).
As before, we can determine $\gamma$ with 3-$\sigma$ precision if $50^{\circ} 
\lesssim \gamma \lesssim 170^o$.  
This range of $\gamma$ covers substantial part of $\gamma$ that is allowed 
in Eq.~(4).
The uncertainty in $\gamma$ is about 
$5^{\circ} - 20^{\circ}$ in this range of $\gamma$, and we achieve a better
determination  of $\gamma$ from the self-tagging $B\rightarrow D K^*$ 
decay modes.

In conclusion, we considered a new method to extract the weak phase $\gamma$
using the triangles shown in Fig.~1.  If one has $3\times 10^8$ $B\bar{B}$ 
pairs at the $\Upsilon (4S)$ resonance, or $10^{11}$ $B$'s at hadron colliders,
one can determine $\gamma$ with 3-$\sigma$ precision or better for 
$80^o (50^{\circ}) \lesssim \gamma \lesssim 150^o (170^o)$ from $B 
\rightarrow D K^{(*)}$ decays that are not Cabibbo suppressed.  
This range of $\gamma$ covers substantial part of $\gamma$ that is allowed 
in Eq.~(4).
One can also repeat for the $D_2 K^{(*)}$ modes instead of $D_1 K^{(*)}$ 
modes, which will provide independent informations on $\gamma$.  For smaller 
$\gamma$, the uncertainty in $\gamma$ is so large that our method is no 
longer useful in extracting $\gamma$. Also our method fails if there is 
no/little strong phase difference $\delta$. One has to resort to some other 
methods in these  cases.  

\vspace{.5in}

{\sl Note Added}

While we were finishing this paper, we received the paper by M. Gronau
and J.L. Rosner \cite{gr}, 
who arrive at the same conclusion as our work.

\acknowledgements
We are grateful to M. Gronau and J.L. Rosner for their correspondence, 
suggestions and encouragements.
This work is supported in part by KOSEF  Contract No. 971-0201-002-2,
through Center for Theoretical Physics at Seoul National University, 
by the Ministry of Education through the Basic Science Research Institute,
Contract No. BSRI-98-2418 (PK).


\begin{table}
\caption{The tagging and detection efficiencies ($f$'s and $\epsilon$'s) 
assumed to estimate the $\Delta \gamma$. See the text for the details.
Numerical values  are for hadron collider ( $\Upsilon (4S)$ ).} 
\label{table1}
\begin{tabular}{cccc}
particles & tagging modes & observation rate ( $f$ )& efficiency ( $\epsilon$ )
\\     \tableline
$K^0, \bar{K}^0$ & $K_S \rightarrow \pi^+ \pi^-$ & 1/3 ( 1/3 ) & 0.1 ( 1.0 )
\\ \tableline
$K^{*0}, \bar{K}^{*0}$ & $K^{*0} \rightarrow K^+ \pi^-$ & 2/3 ( 2/3 ) & 
0.1 ( 1.0 )
\\ \tableline
$K^{*\pm}$ & $K^{*+} \rightarrow K^0 \pi^+$ & 5/9 ( 5/9 )& 0.1 ( 1.0 )
\\
 & $K^{*+} \rightarrow K^+ \pi^0$ & &
\\ \tableline
$D^0, \bar{D}^0$ & $D^0 \rightarrow K^- \pi^+$ & 0.25 ( 0.25 )& 0.1 ( 1.0 )
\\
 & $D^0 \rightarrow K^- \pi^+ \pi^+ \pi^-$ &  & 
\\
 & $D^0 \rightarrow K^- \pi^+ \pi^0 $ &  & 
\\ \tableline
$D^{\pm}$ & $D^+ \rightarrow K^- \pi^+ \pi^+$ & 0.91 ( 0.91 )& 0.1 ( 1.0 )
\\ \tableline
$D_1$ & $D^0 \rightarrow \pi^+ \pi^-$ & $5.85 \times 10^{-3}
( 5 \times 10^{-2} )$ & 0.1 ( 1.0 )
\\
 & $D^0 \rightarrow K^+ K^-$ & &
\end{tabular}
\end{table}

\begin{figure}[h]
\centerline{\epsffile{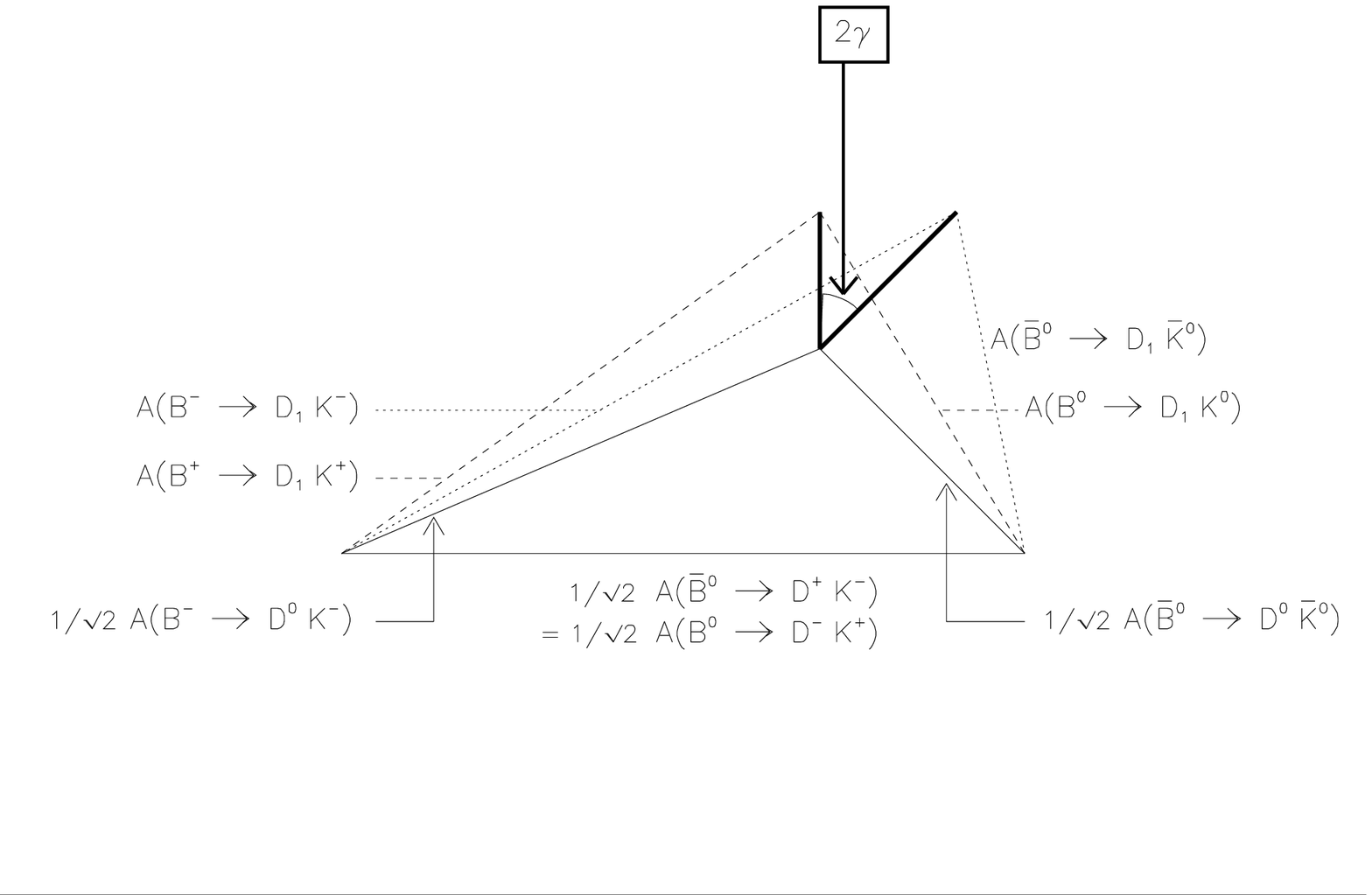}}
\caption{ Three triangles constructed from various $B\rightarrow DK$ modes}
\label{fig1}
\end{figure}

%
%

\begin{figure}[h]
\centerline{\epsffile{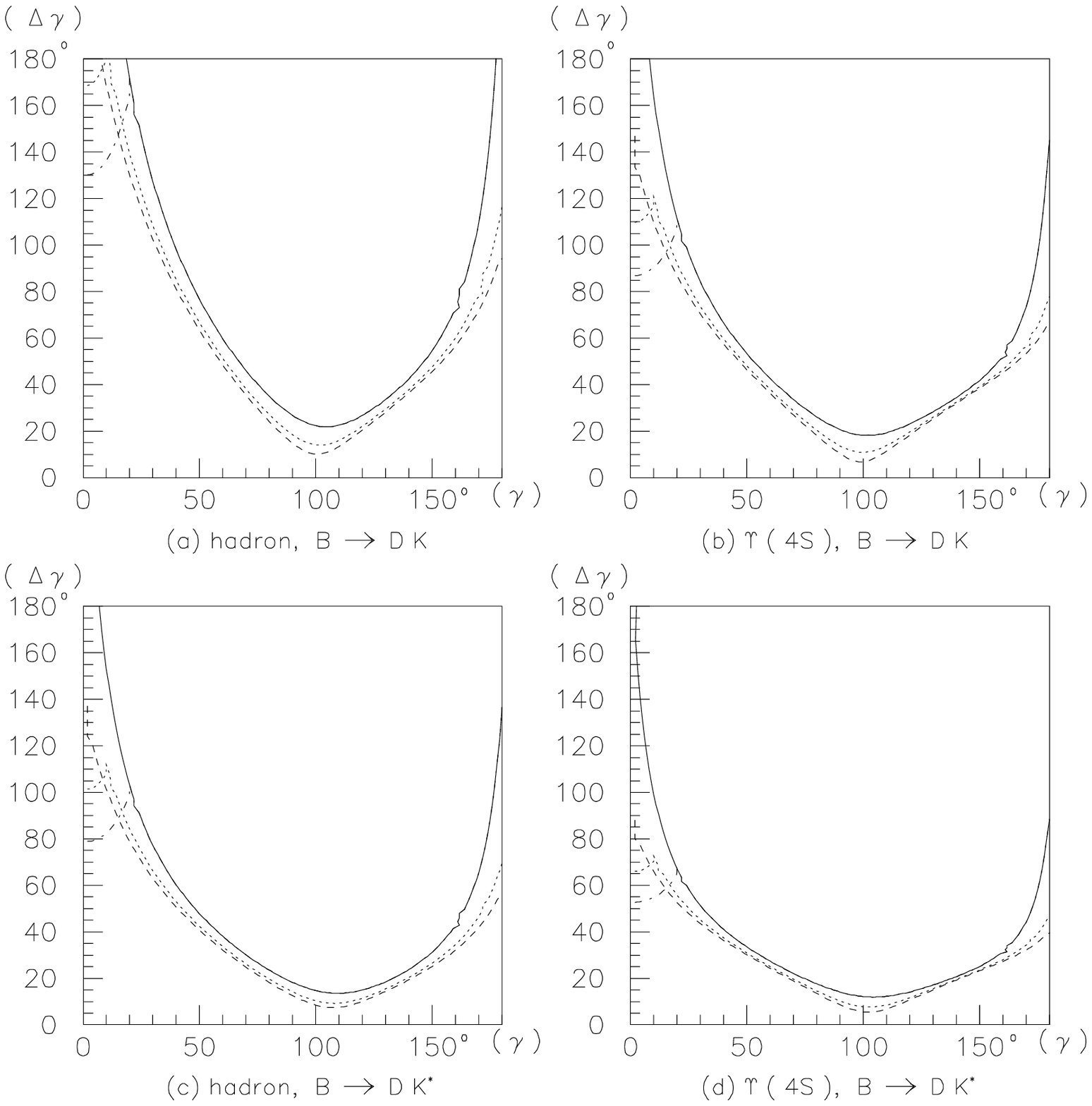}}
\caption{The $\Delta \gamma$ ( error ) plot in the determination of $\gamma$
assuming $10^{11}$ $B$'s at the hadron machines and 
$3 \times 10^{8}$ $B$'s at $B$ factories :
(a) hadron machines and $B \rightarrow D K$,
(b) $B$ factories and $B \rightarrow D K$,
(c) hadron machines and $B \rightarrow D K^*$
and (d) $B$ factories and $B \rightarrow D K^*$.
( $\delta_C = 10^o$ fixed,
real line : $\delta_{C^{'}} = -2 \delta_C$, 
dashed line : $\delta_{C^{'}} =0$,
dotted line : $\delta_{C^{'}} = \delta_C$, 
dashed-dotted line : $\delta_{C^{'}}= 2 \delta_C$.)
}
\label{fig2}
\end{figure}

\begin{references}
\bibitem{quinn} For a review, see the article of 
H. Quinn in {\it Particle Data Book}, R.M. Barnett {\it et al.}, 
Phys. Rev. {\bf D 54}, 507 (1996). 

\bibitem{km} 
M. Kobayashi and T. Maskawa, Prog. Theor. Phys. {\bf 49}, 652 (1973).

\bibitem{aliburas} 
A. J. Buras, TUM-HEP-299-97, hep-ph/9711217 ; 
J.L. Rosner, Nucl. Instrum. Meth. {\bf A 408}, 308 (1998) ;
A. Ali, hep-ph/9801270.

\bibitem{cleo} R. Godang {it et al.}, Phys. Rev. Lett. {\bf 80},
3456 (1998); J. G. Smith, COLO-HEP-395, hep-ph/9803028 ; J. Roy,
Talk at ICHEP98 at Vancouver, Canada.  

\bibitem{GL90} M. Gronau and D. London, Phys. Rev. Lett. {\bf 65},
3381 (1990).

\bibitem{gamma} 
M. Gronau and D. London, Phys. Lett. {\bf B253}, 483 (1991) ;
M. Gronau and D. Wyler, Phys. Lett. {\bf B265}, 172 (1991).

\bibitem{dunietz} 
I. Dunietz, Phys. Lett. {\bf B 270}, 75 (1991).

\bibitem{soni} 
D. Atwood, I. Dunietz and A. Soni, Phys. Rev. Lett. {\bf 78}, 3257 (1997).

\bibitem{gronau98} M. Gronau, CALT-68-2159, hep-ph/9802315.

\bibitem{alilu}
A. Ali, G. Kramer and C.-D. L\"{u}, hep-ph/9804363. 

\bibitem{cleo98} 
M. Athanas {\it et al.}, Cornell University Report CLNS 98--1541 (1998).



\bibitem{rosner} 
M. Gronau and J.L. Rosner, Phys. Rev. Lett. {\bf 79}, 4333 (1997).

\bibitem{neubert} 
M. Neubert and Stech, CERN-TH/97-99, HD-THEP-97-23, hep-ph/9705292.

\bibitem{gr}
M. Gronau and J.L. Rosner, EFI-98-29, FERMILAB-PUB-98-227-T, 
hep-ph/9807447.

\end{references}
\end{document}